\documentstyle[11pt]{article}
\title{\vspace{-1.5in} 
 \hfill {\small{\it Naturam expelles furca}},\\
\hfill{\small {\it  tamen usque recurret...}} \\
\hfill{\small Horatius, {\it Epistolae} {\bf I, 10}, 24 (20 B.C.)} \\~\\~\\
THE WEAK SCALE: \\  DYNAMICAL DETERMINATION \\
{\it VERSUS} \\ ACCIDENTAL STABILIZATION}
\author{ George Triantaphyllou
\thanks{Essay based on talks given at the HEP 2001 Conference 
(Heraklion, Crete) in April 2001 and at the Corfu Summer Institute on 
Elementary Particle Physics (Greece) in September 2001.}
$\;$\\~  \\{\it gtriant$@$telesis.gr}}
\begin{document}
\setlength{\baselineskip}{24pt}
\maketitle
\thispagestyle{empty}
\begin{abstract}
Is it a mere accident that the weak scale is exactly so much smaller than the 
Planck scale, and at the same time exactly so much larger than the QCD scale? 
Or are the experimentally-measured values of the corresponding gauge 
couplings enough to help us determine dynamically these energy scales? And if 
nature has indeed offered us the possibility of such a determination, 
why dismiss it and fix these scales instead by means of arbitrary parameters 
within a multitude of jejune theoretical frameworks which make this wonderful 
hierarchy seem fortuitous? 

\vspace{2.in}
\end{abstract}
\vfill
\setcounter{page}{0}
\pagebreak

The electroweak symmetry is found to be broken at a scale $\Lambda_{K} \sim$ 
1 TeV, roughly three orders of magnitude higher than the QCD scale 
$\Lambda_{C}$ and sixteen orders of magnitude lower than the Planck scale 
$\Lambda_{Pl}$. Has nature already given us ways to really understand these 
large energy-scale ratios, or should on find recourse to new symmetries or 
mechanisms which require fixing these hierarchies by hand? Is the hierarchy 
problem a mere question of {\it ad hoc} weak-scale stabilization as current
literature seems to suggest, or more a question of weak-scale dynamical 
determination?

An order-of-magnitude relationship can be conjectured for $\Lambda_{K}$, 
since 
\begin{equation}
\Lambda_{K} \approx \Lambda_{GUT} e^{- \alpha_{L}^{-1}}
\label{eq:equa}
\end{equation}

\noindent where $\alpha_{L}$ is the weak gauge coupling and $\Lambda_{GUT}$
the gauge-coupling unification scale which is usually found to be a couple
of orders of magnitude smaller than $\Lambda_{Pl}$. 
However, arguments supporting such a 
relation were hitherto limited to mere numerology or dimensional analysis, 
and no consistent theory could accommodate it. Could it be just a happy 
coincidence that the actual values of the quantities appearing in 
Eq.\ref{eq:equa} render this relationship plausible?

Not likely. Katoptron theory \cite{george1}-\cite{george5} which was 
developed during the last three years incorporates naturally the above 
relationship with its gauge-coupling unification including the 
katoptron-group coupling \cite{george2}-\cite{george2a}, 
providing us with a solid framework producing uniquely the observed scale 
hierarchies without arbritrarily adjustable parameters which would make it 
look like an accident - based only on anthropic-principle arguments perhaps - 
that the weak scale is not 5 or ``pick-your-own-number" orders of magnitude 
larger than what is measured experimentally. 

How does katoptron theory determine dynamically these scales? It 
introduces new fermions called katoptrons, having dynamical (``constituent") 
masses on the order of the weak scale and mirror weak quantum-number 
assignments \cite{george1}, along with a gauged katoptron-family ``horizontal" 
symmetry (denoted by $SU(3)^{\prime}$ below) corresponding to an interaction 
which becomes strong at $\Lambda_{K}$. Non-perturbative dynamics near 1 TeV 
render the {\it ad unguem} study of the corresponding phenomenology at lower 
energies difficult, judging also from our QCD experience. Nevertheless, it is 
important to realise that the theory's full gauge-field and matter content 
describing fermions and the interactions between them is, as usual, 
accurately and absolutely well-defined within the representation theory of 
Lie groups \cite{george4}.

To be more precise, under the gauge symmetry
 $SU(3)_{C} \times SU(2)_{L} \times U(1)_{Y} \times
SU(3)^{\prime}$, fermions transform as
\begin{eqnarray}
&{\rm SM\;fermions} &{\rm Katoptrons} \nonumber
\\
&q_{L}:({\bf 3,\;2},\;1/3,\;{\bf 1})_{i}  & q_{R}^{K}:({\bf
3,\;2},\;1/3,\;{\bf 3}) \nonumber \\ &l_{L}:({\bf
1,\;2},\;-1,\;{\bf 1})_{i}  &l_{R}^{K}:({\bf 1,\;2},\;-1,\;{\bf
3})\nonumber \\&q_{R}^{c}:({\bf {\bar
3},\;1},\;~^{-4/3}_{+2/3},\;{\bf 1})_{i}  & q_{L}^{K\;c}: ({\bf
{\bar 3},\;1},\;~^{-4/3}_{+2/3},\;{\bf 3}) \nonumber
\\&l_{R}^{c}:
 ({\bf 1,\;1},\;~^{0}_{2},\;{\bf 1})_{i} & l_{L}^{K\;c}:
 ({\bf 1,\;1},\;~^{0}_{2},\;{\bf 3}),
   \nonumber \\
\end{eqnarray}
\noindent where $q$ and $l$ denote quarks and leptons respectively
and $i=1,2,3$ is a Standard-model-generation index. 

Lee and Yang are the first recorded to have presciently speculated on the 
existence of mirror fermions in the {\it locus classicus} on parity violation 
in 1956, since such fermions would restore a frequently-unheeded symmetry of 
nature, nearly as essential as the matter-antimatter symmetry. After that 
work, several authors used mirror fermions in their models, but no model 
used strongly-interacting mirror fermions which could 
predict and reproduce correctly the weak scale like katoptron theory does.  

The theory is based on the main concept that strongly-interacting fermions
form condensates which break dynamically the electroweak symmetry, a general 
idea initially thought by Weinberg and Susskind in the late seventies. 
The strongly interacting fermions are katoptrons in the present setting, and 
the source of this interaction is a gauged katoptron-family symmetry which 
becomes non-perturbative around 1 TeV. The fundamental reason for the large 
hierarchy between the weak and the unification scales is therefore to be 
found in the logarithmic renormalization of the katoptron-family gauge 
coupling. It is this family symmetry which prevents the pairing of the 
Standard-Model fermions with their mirror partners, which would otherwise 
give them GUT-scale gauge-invariant masses. The new composite degrees of 
freedom arising at lower energies have to be studied within an 
effective-potential framework valid only up to a cut-off scale around 
$\Lambda_{K}$. 

Theories involving strongly-interacting gauged
family symmetries, broken or not, have previously appeared, but none involves 
mirror fermions with masses around the weak scale. Furthermore, models 
introducing new strongly-interacting fermions for dynamical electroweak 
symmetry breaking, usually under the name of technicolor and too numerous to
cite here, proved to have fatal flaws, exactly because technifermions have, 
contrary to katoptrons, Standard-Model-type weak quantum-number assignments.

A host of problems plaguing technicolor models are solved within katoptron
theory.  First of all, all symmetries are gauged and if broken, they break
around 1 TeV. Therefore, there is no need for fine-tuning of parameters in 
order to produce reasonable fermion masses, and very light pseudo-goldstone 
bosons are not expected. Moreover, there are no flavour-changing neutral 
currents involving Standard-Model fermions, forasmuch as these fermions are 
not included in the same group representations together with katoptrons, 
unlike what happens in extended-technicolor (ETC) models. In addition, the 
absence of very light degrees of freedom extricates the theory from problems 
related to anomaly-matching conditions. Furthermore, the self-breaking of the 
katoptron group, discussed {\it in extenso} in \cite{george5}, 
renders the existence of any light elementary scalar particles obsolete. 
One should finally note that gauge-invariance prohibits the appearance of 
certain dimension-six operators in the effective Lagrangian which would 
otherwise produce indirectly observable new-physics effects and create what 
is known as ``the LEP paradox".

A new set of experimental data confuting technicolor-like models appeared by 
the beginning of the nineties, with LEP results indicating that the 
$\Delta\rho$ and $S$ parameters were consistent with zero, their 
Standard-Model values. The katoptron
theory, albeit it bears a {\it prima facie} propinquity with such models, is 
able to circumvent both of these problems. The $\Delta\rho$ parameter is 
naturally small within this framework, since the operators feeding masses 
to the top and bottom quarks are electroweak invariant, unlike the situation 
in ETC models, and therefore cannot influence considerably the masses of the 
$SU(2)_{L}$ gauge bosons. 

In addition, the $S$ parameter in katoptron theory can be small for three
different reasons counterpoising the introduction of new fermion flavors. 
First, the strong gauge group is broken, reducing the expected contribution 
roughly by a factor of two. Second, Majorana neutrinos appearing naturally 
in the theory can give negative contributions to $S$. Since these corrections 
are larger for lighter katoptron neutrinos, a small $S$ parameter could 
indicate that the lightest katoptron neutrino mass lies just beyond present 
experimental bounds of about 100 GeV \cite{george2}. And, last but not least, 
anomalous right-handed weak couplings $\delta g_{R}^{t,b}$ of the top and 
bottom quarks can cause large deviations of the quark asymmetries $A_{t,b}$ 
from their Standard-Model values and therefore produce negative 
contributions to $S$ large enough to render this parameter consistent with
zero. Note moreover that $S$ is not constrained severely enough to allow the 
characterisation of such a cancellation as ``fine tuning"
\footnote{Instances where correct interpretations of physical phenomena are
arrogantly and yet erroneously dismissed in favour of less natural ones 
because an expected deviation of a crucial parameter is ostensibly missing 
abound and date back to antiquity. The apparent absence of star parallax 
during the Earth's revolution around the sun, to recall a classic example, 
led to the dogmatic rejection of the correct heliocentric planetary system 
developed by Aristarchos of Samos (310-230 B.C.) in favour of Aristotle's 
geocentric system for more than a millenium.}
\cite{george1}.

Interestingly enough, katoptron theory gives rise to operators 
able to produce the current 2.6 $\sigma$ deviation of the asymmetry $A_{b}$, 
not to be confused with $R_{b}$, from its Standard-Model value, contrary to 
other new-physics models. Since these operators are proportional to the 
relevant quark masses \cite{george5}, the corresponding asymmetry for the 
top quark $A_{t}$ could be much larger than the one predicted by the 
Standard Model and could be {\it ipso facto} directly related to the small 
value of the $S$ parameter. Data on the top-quark weak couplings from the 
next leptonic collider (the NLC/JLC or TESLA for instance) will be quite 
enlightening at this point. Moreover, katoptron theory can also be made 
consistent with the current deviation of the experimentally-measured 
anomalous muon magnetic moment from its Standard-Model value, but this is 
not unique to the theory inasmuch as other new-physics models can also 
produce this effect. 

Furthermore, the general fermion mass matrix involving both fermions and
katoptrons hints to a solution of the strong CP problem, unlike other
dynamical symmetry-breaking models which are usually adding their strong CP 
problems to the QCD one. The various gauge dynamics in katoptron theory 
should be the ones which determine the entries of this mass matrix, and not 
the arbitrary Yukawa couplings 
appearing in  other ``new-physics" models. These entries should finally 
determine not only the masses of all the fermions of the theory, but also 
their mixings with each other within a generalized fermion-mixing matrix. 
In particular, the heaviness of the top quark should lead to a small 
$|V_{tb}|$ mixing entry equal to 0.95 \cite{george1}, \cite{george5} 
{\it grosso modo}, another challenging measurement for the next colliders.

The embedding of katoptron theory within a more fundamental unified gauge 
theory at the Planck scale, the $E_{8} \times E_{8}^{\prime}$ supersymmetric 
gauge theory in 10 dimensions of Green and Schwarz \cite{george4}, provides 
the first example of a coset-space dimensional reduction process leading 
to the Standard Model plus a new strongly-interacting fermion sector breaking 
dynamically the electroweak symmetry at the right scale without the need of 
any light elementary scalar fields. It is in parallel the first example of a 
unified dynamical electroweak-symmetry breaking framework which,  
conceptually simple and succinctly defined though it is, approaches 
convincingly the Planck scale and the quantum-gravity dynamics associated 
with it, a {\it sine qua non} at least for the verisimilitude of a modern 
theory of elementary-particle physics.

The breaking of $E_{8} \times E_{8}^{\prime}$  down to a diagonal subgroup 
at the unification scale leaves no mysterious ``shadow" or ``hidden" world 
below the compactification scale left behind, apart of course from the one 
around 1 TeV waiting to be discovered in the next generation of high-energy 
experiments. Furthermore, this embedding renders the theory anomaly-free, 
something not clear when one considers only the lighter degrees of freedom 
of katoptron theory. In addition, the compactification process sheds light 
{\it inter alia} on the origins of supersymmetry breaking, parity violation 
and the number of fermion generations.

The setting described above confirms Georgi $\&$ Glashow's ``desert 
hypothesis". Moreover, it shows that the deduction that there already exists 
an unambiguous sign of low-energy supersymmetry given that gauge couplings 
unify at energies close to the Planck scale {\it non sequitur}. 
It renders also apparent that the katoptron-theory's solution to the 
hierarchy problem obviates the need of low-energy supersymmetry, since it is 
a gauge and not a space-time symmetry which stabilises the weak scale. Since 
one obtains here also coupling unification, one still has a ``prediction" of 
$\sin^{2}\theta_{W}$, which is anyway a contrived rephrasing of the 
elementary fact of two-dimensional Euclidean geometry that a single 
parameter is enough to force three non-parallel lines to cross at the same 
point. 

In the katoptron-theory framework, supersymmetry breaks at the 
unification scale, leaving itself and the appertaining suppositious 
superpertners (and the cosmologically-dangerous dilaton if one wishes to  
include gravity in the discussion) at scales much higher than the ones 
accessible in the experiments to come. The superpartners are replaced by 
katoptron fields with masses around 1 TeV. The large unification scale is 
consistent with small neutrino masses and their mixings \cite{george2}, and 
naturally suppresses proton decay without imposing any new symmetry or otiose 
mechanism for that purpose. However, the predicted proton-lifetime lies just 
beyond present experimental bounds, and proton-decay events could be 
observed in the near future.

After this discussion, low-energy supersymmetric models {\it ex cathedra}, 
incontrovertibly popular and plenteously cited though they might be, seem to 
be rather specious and merely self-consistent with regards to the relation 
of the weak scale to the top-quark mass, but are however unable to determine 
how exactly and at which scale supersymmetry breaks without cyclical 
arguments and without adding new parameters and ``hidden sectors" by hand. 
It seems consequently quite likely that the accidental and by no means unique 
gauge-coupling unification in models with low-energy supersymmetry is just 
an indication that the general concept of group unification and the 
``desert hypothesis" of GUTs is {\it lato sensu} correct. 

This concept does not form the basis of many other attempts to solve 
the hierarchy problem, and constitutes admitedly a considerable improvement 
of our understanding of nature. Beyond that nonetheless, and with regards to 
which type of particles one might expect around 1 TeV, it might well prove to 
be one of the most blatant ``red herring" cases in the history of Physics. 
More generally, the quest for typical Planck-scale signatures like 
supersymmetry and extra dimensions in the next HEP experiments may just be a
vain pursuit of an {\it ignis fatuus}.

To conclude, we have described in brief the perspicuous advantages of
katoptron theory, which is unique in addressing succesfully the essence of 
the hierarchy problem, {\it scilicet} not only why the weak scale is stable 
but also why it is thirteen orders of magnitude smaller than the unification 
scale and not some other arbitrarily adjustable number, a {\it non dit} of 
- inherently scale-blind - supersymmetric and recent large-extra-dimensions 
models for instance. Furthermore, the theory is consistent with the current 
2.6 $\sigma$ deviation of the asymmetry $A_{b}$ oppugning the Standard Model. 

Important theoretical issues which need further inquiry include the
- order-of-magnitude at least - calculation of the mass-matrix entries from 
first principles, and the physical criteria which lead to the specific 
compactification manifold producing this theory. From the phenomenological
point of view, the new rich structure introduced should lead to a detailed
study of the expected signals in the next set of experiments. Obviously,
if no significant deviations of the parameters $|V_{tb}|$ and $A_{t}$ from
their Standard-Model values are measured, and if no new mirror fermions
with masses below about 300 GeV are found, this theory will be
most likely excluded.

On the other hand, the discovery 
of new heavy fermion flavours at the LHC, maybe bound in scalar states, and 
the subsequent measurement of their precise chirality at a very powerful 
and adequately optimized future lepton collider 
(a muon collider or the CLIC for instance) reaching energies where katoptron 
dynamics become perturbative \cite{george3} might - as the final, unbiased
and unique judge - provide tangible proof of katoptron theory and settle 
{\it a fortiori} this quite important issue before high-energy physics 
confronts the ineluctable desert. Nearly four decades after the impressive 
triumph of strong gauge dynamics lead by the QCD ``flagship" over pleasant 
formalistic excursions, it remains to be seen whether history repeats itself 
{\it mutatis mutandis}.

\noindent {\bf Acknowledgements} \\
We thank the organisers of the conferences for giving us the opportunity to 
communicate our views, M. Axenides, P. Ditsas, M. Kaku, E. Kiritsis and 
G. Senjanovic for interesting discussions, and the physicists cited in the 
first five original papers on katoptron theory for some useful tools and 
ideas.

\end{document}